\documentclass[11pt,a4paper]{article}
\usepackage{amsmath}

\usepackage[psamsfonts]{amssymb}
\usepackage{amsmath}
\usepackage{epsfig}

\author{H. Mohseni Sadjadi\footnote{mohseni@phymail.ut.ac.ir}
\\ {\small Department of physics, University of Tehran ,}
\\ {\small P.O.B. 14395-547, Tehran 14399-55961, Iran}}
\title{ Generalized second law in modified theory of gravity}

\begin{document}
\maketitle
\begin{abstract}
In the context of modified theory of gravity ($f(R)$ gravity) we
try to study the conditions needed for validity of the generalized
second law.

\end{abstract}

\section{Introduction}
Recently, motivated by astrophysical data which indicate that the
expansion of the universe is accelerating \cite{acc}, the modified
theory of gravity (or $f(R)$ gravity) which can explain the
present acceleration without introducing dark energy, has received
intense attention \cite{mod}.

Determining thermodynamic parameters of an (accelerated) expanding
universe and verification of the first and the second law for
different cosmological horizons \cite{law}; investigating the
relation between dynamics and thermodynamics of the universe
\cite{jacob};  studying the conditions required for validity of
the generalized second law (GSL) \cite{GSL}, and so on, have also
been the subjects of many researches in recent years.

In the modified theory of gravity, instead of Friedman equations
we must utilize modified Friedmann equations which may include the
powers of Ricci scalar as well as its time derivatives. Besides,
the relation of the entropy to the area of the horizon is also
different with Einstein theory of gravity. So it is of interest to
see how, in the framework of $f(R)$ gravity, thermodynamic
properties of the universe may be modified.

In this paper we try to find necessary conditions for validity of
GSL in the framework of $f(R)$ gravity in Friedmann Robertson
Walker (FRW) universe. The future event horizon is taken as the
horizon of the universe and the temperature is supposed to be
proportional to the Gibbons-Hawking temperature \cite{hawk}. The
range of the proportionality constant will be determined through
some examples. Using the relation between the entropy assigned to
the horizon and its area derived from the Noether charge method,
we study the behavior of horizon entropy with respect to the
comoving time. We assume that the horizon is in thermal
equilibrium with its environment which is filled with perfect
fluids. Then from modified Friedmann equations and the first law
of thermodynamics, the time derivative of the fluid entropy and
subsequently the time derivative of the total entropy is
determined. At the end we elucidate our results via two examples.

We use the units $\hbar=c=G=k_{B}=1$.

\section{Thermodynamics and GSL in FRW universe in modified gravity}

The action of modified theory of gravity with the inclusion of
matter is given by
\begin{equation}\label{1}
S={1\over 16\pi}\int d^4x\sqrt{-g}f(R)+S_m,
\end{equation}
where $S_m$ is the matter action, $R$ is the Ricci scalar
curvature and $f(R)$ is an arbitrary real function. Variation of
the action with respect to the metric gives
\begin{equation}\label{2}
R_{\mu\nu}f'(R)-{1\over 2}g_{\mu\nu}f(R)+g_{\mu\nu}\Box
f'(R)-\nabla_\mu \nabla_\nu f'=8\pi T^m_{\mu\nu},
\end{equation}
where the prime denotes the derivative with respect to $R$, and
$T^ m_{\mu\nu}$ is the energy-momentum tensor of the matter
fields. For spatially flat FRW metric with scale factor $a(t)$:
\begin{equation}\label{3}
ds^2=-dt^2+a^2(t)(dx^2+dy^2+dz^2),
\end{equation}
Eq.(\ref{2}) yields
\begin{eqnarray}\label{4}
8\pi\rho&=&{f(R)\over 2}-3(\dot{H}+H^2-H{d\over dt})f'(R)
\nonumber \\
8\pi P&=&-{f(R)\over 2}+(\dot{H}+3H^2-{d^2\over dt^2}-2H{d\over
dt})f'(R).
\end{eqnarray}
The Hubble parameter is given by $H=\dot{a}/a$, and the Ricci
scalar is obtained as $R=6\dot{H}+12H^2$. The over dot indicates
the derivative with respect to the comoving time $t$. $\rho$ and
$P$ are the density and the pressure of the matter which behaves
as a perfect fluid at large scale:
\begin{equation}\label{5}
T^m_{\mu \nu}= (\rho+ P )U_{\mu}U_{\nu}+ Pg_{\mu \nu},
\end{equation}
where $U_{\mu}$ is the four velocity of the fluid.

The radius of the future event horizon, $R_h$, is given by
\begin{equation}\label{6}
R_h(t)=a(t)\int_t^{\infty}\frac{dt'}{a(t')}.
\end{equation}
Note that the event horizon exists, $R_h(t)\in \Re$, when the
above integral converges, i.e., $\int_t^\infty dt'/a(t')\in
\Re^{+}$. If at a time denoted by $t_s$, the Big Rip singularity
occurs, we must replace $\infty$ by $t_s$ in the integration.
Using Eq.(\ref{6}), we can verify that $R_h$ satisfies the
following equation
\begin{equation}\label{7}
\dot{R_h}=HR_h-1.
\end{equation}

In the context of Einstein theory of gravity, the entropy of a
black hole is given by Bekenstein-Hawking relation \cite{bek}
\begin{equation}\label{8}
S_h= {A\over 4},
\end{equation}
where $A$ is the area of the event horizon. In the same way one
can assign an entropy to the cosmological future event horizon
whose area is $A_h=4\pi R_h^2$. This entropy which is given by
$S_h={A_h\over 4}$, may be regarded as a measure of information
hidden behind the horizon. In de Sitter space-time $R_H={1\over
H}$ and the future event horizon becomes the same as the de Sitter
(Hubble) horizon. In this space time the temperature, which is
dubbed as Gibbon Hawking temperature \cite{hawk},  can be
determined in terms of horizon radius as $T={H\over 2\pi}$.

In $f(R)$ gravity, Noether charge method can be used to obtain the
horizon entropy \cite{noe}
\begin{equation}\label{9}
S_h={1\over 4}\int_A f'(R)dA,
\end{equation}
where the integration is taken over the surface of the horizon,
$A$. So in FRW universe, where the scalar curvature is spatially
constant, the entropy is obtained as
\begin{equation}\label{10}
S_h={Af'(R)\over 4}.
\end{equation}

In the following we choose the future event horizon as the horizon
of the universe. Differentiating Eq. (\ref{10}) with respect to
the comoving time gives
\begin{equation}\label{11}
\dot{S_h}=2\pi \dot{R_h}R_hF+\pi R_h^2\dot{F}.
\end{equation}
We have defined $F=f'(R)$. In Einstein theory of gravity the above
equation reduces to $\dot{S_h}=2\pi R_h\dot{R_h}$. Note that in a
super-accelerated universe defined by $\dot{H}>0$, $R_h$ is
decreasing: $\dot{R_h}<0$ \cite{sad}, therefore $\dot{S_h}<0$. But
in modified theory this is not the case and, depending on the
function $f(R)$, one may have $\dot{S_h}>0$. Consider the model
$f(R)=\alpha R^m$, $\alpha, m \in \Re$. For $\dot{H}>0$ we have
$R>0$, hence $\dot{S_h}\geq 0$ leads to
\begin{equation}\label{12}
\alpha\left({\dot{R_h}\over R_h}+{(m-1)\dot{R}\over 2R}\right)\geq
0.
\end{equation}
It is clear that for $\alpha=m=1$ the above inequality cannot be
satisfied in a super-accelerated universe.

If $S_{in}$ is the entropy of the matter inside the horizon, then
the first law of thermodynamics states
\begin{equation}\label{13}
TdS_{in}=dE+PdV=(P+\rho)dV+Vd\rho.
\end{equation}
By taking $V={4\over 3}\pi R_h^3$, we arrive at
\begin{equation}\label{14}
T\dot{S_{in}}=4\pi(P+\rho)R_h^2\dot{R_h}+{4\over 3}\pi
R_h^3\dot{\rho}.
\end{equation}
From Eq.(\ref{7}), and energy conservation relation
\begin{equation}\label{15}
\dot{\rho}+3H(P+\rho)=0,
\end{equation}
we can write Eq.(\ref{14}) in the form
\begin{equation}\label{16}
T\dot{S_{in}}=-4\pi(P+\rho)R_h^2.
\end{equation}
$S_{in}$ is a decreasing (increasing) function of time when
$w>(<)-1$, where  $w={P\over \rho}$ is the effective equation of
state (EOS) parameter of the perfect fluid filling the universe.
In the model $f(R)=R$, Eq. (\ref{16}) becomes
$T\dot{S_{in}}=\dot{H}R_h^2$, hence $\dot{S_{in}}>0$ is satisfied
when $\dot{H}>0$. Note that in $f(R)$ theory of gravity we have
\begin{equation}\label{17}
w=-1+{-4\dot{H}F+2H\dot{F}-2\ddot{F}\over
f(R)-6(\dot{H}+H^2-H{d\over dt})F}.
\end{equation}
Hence in contrast to Einstein theory of gravity, in $f(R)$ models,
$\dot{H}>0$ does not requires $w<-1$ and $\dot{H}>0$ is not a
necessary condition for $\dot{S_{in}}>0$.

Temperature of the horizon, $T$, which is taken the same as the
fluid temperature, is independent of the gravity theory that leads
to the horizon geometry (see e.g. Ref. \cite{wang}). In $f(R)$
gravity, like Einstein theory of gravity, and in the absence of a
well defined temperature for cosmological horizon, we assume that
$T$ is proportional to Gibbons-Hawking temperature
\begin{equation}\label{18}
T=\frac{bH}{2\pi}.
\end{equation}
$b$ is a real constant. In a de Sitter space we must take $b=1$.
Indeed the parameter $b$ shows the deviation from Gibbons-Hawking
temperature. Inserting the modified Einstein-Friedmann equations
(Eqs. (\ref{4})), and (\ref{18}), into Eq. (\ref{16}), results in
\begin{equation}\label{19}
\dot{S_{in}}={2\pi R_h^2\over bH}(\dot{H}+{1\over 2}{d^2\over
dt^2}-{1\over 2}H{d\over dt})F.
\end{equation}

The total entropy of the universe, denoted by $S$, is the sum of
the matter entropy inside the horizon, $S_{in}$, and $S_h$: $ S=
S_{in}+S_h$. The generalized second law (GSL) states that the
total entropy is not a decreasing function of time:
\begin{equation}\label{20}
\dot{S}=\dot{S_{in}}+\dot{S_{h}}\geq 0.
\end{equation}
Following our previous results, this leads to
\begin{equation}\label{21}
\dot{S}=\pi R_h^2\Big((1-{1\over b})\dot{F}+2({\dot{R_h}\over
R_h}+{\dot{H}\over bH})F+{1\over bH}\ddot{F}\Big)\geq 0,
\end{equation}
which is more complicated with respect to the model $f(R)=R$ in
which Eq. (\ref{21}) reduces to the following simple inequality
\begin{equation}\label{22}
(H^{1\over b}R_h\dot{)}\geq 0.
\end{equation}

To illustrate our results let us consider some examples. As the
first example consider a quasi de Sitter FRW space time, defined
by
\begin{equation}\label{23}
H=H_0+H_0\epsilon t+O(\epsilon^2),\,\,\, \epsilon:={\dot{H}\over
H^2}\ll 1,\,\,\, \dot{\epsilon}=\mathcal{O}(\epsilon^2).
\end{equation}
In this space time the future event horizon is given by
\begin{equation}\label{24}
R_h={1\over H}(1-{\dot{H}\over H^2})+\mathcal{O}(\epsilon^2),
\end{equation}
leading to
\begin{equation}\label{25}
\dot{R_h}\simeq -{\dot{H}\over H^2}.
\end{equation}
Therefore ${\dot{R_h}\over R_h}\simeq -{\dot{H}\over H}$ and
$\dot{S}\geq 0$ reduces to
\begin{equation}\label{26}
\dot{S}\simeq \pi R_h^2\left(1-{1\over
b}\right)(\dot{F}-{2\dot{H}\over H}F)\gtrapprox 0.
\end{equation}
Consider the model
\begin{equation}\label{27}
f=\beta R+\alpha R^m,\,\, \beta, \alpha , m \in \Re.
\end{equation}
By using $\dot{R}\simeq 24H\dot{H}$ and $\ddot{R}\simeq
24\dot{H}^2$, one can verify that the GSL is satisfied up to the
order $\mathcal{O}(\epsilon^2)$, provided that
\begin{equation}\label{28}
(1-{1\over b})\dot{H}\Big(\alpha m(m-2)(12H^2)^{m-1}-\beta
\Big)\gtrapprox 0.
\end{equation}
For $\alpha=0$ and $\beta=1$, corresponding to Einstein theory of
gravity, the above equation reduces to
\begin{equation}\label{29}
-(1-{1\over b}){\dot{H}\over H}\gtrapprox0,
\end{equation}
showing that for a super-accelerated universe ($\dot{H}>0$),
$\dot{S}\gtrapprox 0$ is satisfied if $b\lessapprox 1$ and for
quintessence phase ($\dot{H}<0$), GSL is respected when
$b\gtrapprox 1$.

If $\beta=0$ and $\alpha>0$, $\dot{S}\gtrapprox 0$ is satisfied
when $\alpha m(m-2) (1-{1\over b})\dot{H}\gtrapprox 0$. So if
${\dot H}>0$ and $m>2$, we must have $b\gtrapprox 1$. Therefore in
modified gravity in contrast to the Einstein theory of gravity, a
super-accelerated FRW model, which departs slightly from de Sitter
space, may have temperature greater than Gibbons-Hawking
temperature and meanwhile respects the GSL.

As another example consider again the $f(R)$ gravity model defined
by Eq. (\ref{27}). Assume \cite{examp}
\begin{equation}\label{30}
a=a_0(t_s-t)^{-n}, n>0.
\end{equation}
In this model
\begin{equation}\label{31}
H={n\over {t_s-t}}, \,\,\, R={{12n^2+6n}\over (t_s-t)^2}.
\end{equation}
Hence Eq. (\ref{30}) describes a super accelerated FRW universe,
$\dot{H}>0$, with a Big Rip singularity at $t=t_s$. The future
event horizon radius is
\begin{equation}\label{32}
R_h={{t_s-t}\over {n+1}}.
\end{equation}
To show that Eq. (\ref{30}) may be a solution of Eq. (\ref{4}) in
$f(R)$ gravity, one can assume that the perfect fluid inside the
event horizon includes two components with EOS: $P_1=\gamma_1
\rho_1$ and $P_2=\gamma_2\rho_2$, where the constant $\gamma$'s
are the EOS parameters.  These components satisfy
\begin{eqnarray}\label{33}
8\pi\rho_1&=&{\beta R\over 2}-3\beta(\dot{H}+H^2)
\nonumber \\
8\pi P_1&=&-{\beta R\over 2}+\beta(\dot{H}+3H^2),
\end{eqnarray}
and
\begin{eqnarray}\label{34}
8\pi\rho_2&=&{\alpha R^m\over 2}-3m\alpha(\dot{H}+H^2-H{d\over
dt})R^{m-1}
\nonumber \\
8\pi P_2&=&-{\alpha R^m\over 2}+m\alpha(\dot{H}+3H^2-{d^2\over
dt^2}-2H{d\over dt})R^{m-1}.
\end{eqnarray}
Besides, each component ($i=1,2$) satisfies the energy
conservation equation
\begin{equation}\label{35}
\dot{\rho_i}+3H(\gamma_i+1)\rho_i=0.
\end{equation}

Using Eqs. (\ref{33}), (\ref{34}) and Eq. (\ref{35}) and after
some calculations it may be verified that Eq. (\ref{4}) is
satisfied in the case (\ref{27}) provided that:
\begin{equation}\label{36}
{1\over 1+\gamma_1}={m\over 1+\gamma_2}=-{3n\over 2},
\end{equation}
which results in $\gamma_1<0$, and $\gamma_2<(>)-1$ if $m>(<)0$.
Using Eq. (\ref{21}), one can verify that in this model GSL is
respected only for times satisfying
\begin{eqnarray}\label{37}
&&-m\alpha\Big((m-2)(b-1)n+2m^2+1-3m\Big)\Big(6n(1+2n)\Big)^{m-1}\times\nonumber
\\
&&(t_s-t)^{2m-2}+\beta n(b-1)\leq 0.
\end{eqnarray}
In Einstein theory of gravity ($\alpha=0, \beta=1$), $\dot{S}>0$
is satisfied only when $b< 1$, in this case $\dot{S}>0$ holds
$\forall t<t_s$. In the modified theory, depending on the values
of $m,n,\alpha$, and $\beta$, $\dot{S}>0$ holds only for times
belonging to special domain specified by the Eq. (\ref{37}) and we
may have $\dot{S}>0$ while $b>1$. We can also restrict the values
of the parameters to specific domains such that GSL holds for
$\forall t<t_s$, e.g., for
$m\alpha\left[(m-2)(b-1)n+2m^2+1-3m\right]>0$, GSL holds $\forall
t<t_s$ provided that $\beta(b-1)<0$.

It seems that GSL does not hold near the Big Rip singularity,
$t\simeq t_s$, for $\beta(b-1)>0$. This may be related to the fact
that in our classical computation we have ignored the contribution
of the radiation energy density, generated by semiclassical
particle creation from Rindler horizon near the Big Rip
\cite{linder}, in the total entropy.

For an adiabatic expansion ($\dot{S}=0$), the following equations
hold:
\begin{eqnarray}\label{38}
&&-m\alpha\Big((m-2)(b-1)n+2m^2+1-3m\Big)\Big(6n(1+2n)\Big)^{m-1}=0,
\nonumber \\ && \beta n(b-1)=0,
\end{eqnarray}
which in Einstein theory of gravity ($\beta=1, \alpha=0$), infers
$b=1$. Therefore, in this theory,  during an adiabatic expansion
the temperature is the same as Gibbons-Hawking temperature. While
in the modified theory of gravity ($\alpha\neq 0$), in order to
have $\dot{S}=0$, $(m-2)(b-1)n+2m^2+1-3m=0$ and $\beta(b-1)=0$
must be satisfied. Thereby $\beta\neq 0$ leads to $b=1$ and
$m=1/2$; and $\beta=0$ implies $b=1+{(1-m)(2m-1)\over n(m-2)}$, so
depending on the values of $m$ and $n$, in adiabatic expansion,
the temperature may be less or more than the Gibbons-Hawking
temperature.

\end{document}